\documentclass[amsmath]{astronomycomm}

\usepackage{graphicx}
\usepackage{amssymb}
\usepackage{bm}

\begin{document}

\title{Probing Nearby Electron Sources and Local Diffusion with the TeV All-Electron Spectrum}

\shorttitle{Nearby Electron Sources and Local Diffusion}
\shortauthors{Xie and Yang}

\author{Zhen Xie}
\affiliation{School of Astronomy and Space Science, University of Science and Technology of China, Hefei, 230026, China}
\email{zhenxie@mail.ustc.edu.cn}

\author{Ruizhi Yang}
\affiliation{School of Astronomy and Space Science, University of Science and Technology of China, Hefei, 230026, China}
\affiliation{TIANFU Cosmic Ray Research Center, Chengdu, Sichuan, People's Republic of China}
\email[show]{yangrz@ustc.edu.cn}

\begin{abstract}
Radiative losses restrict multi-TeV cosmic-ray electrons to young nearby accelerators, allowing the all-electron spectrum (electrons plus positrons) to retain information about particle release and local transport. We investigate how a nearby source can shape this spectrum using Vela as a representative example. We compare a birth-time burst with continuous release for the same total injected electron energy and examine the dependence on the diffusion coefficient, injection index, and source energetics. The burst spectrum develops a cooling-limited boundary near $28~{\rm TeV}$ at the age of Vela, whereas continuous release produces a broader high-energy tail through the contribution of recently injected particles. The flux at Earth is largest over a finite range of diffusion coefficients, reflecting the competition between inefficient arrival and spatial dilution. Harder injection spectra and larger released energies increase the source contribution in the multi-TeV region, and we map the combinations of injection index and released electron energy that can produce an observable spectral feature. For anisotropic diffusion, the source visibility also depends on the angle between the Vela--Earth direction and the local ordered magnetic field. Future direct measurements extending the all-electron spectrum beyond several TeV could therefore distinguish different release histories and constrain both the energetics of nearby sources and local cosmic-ray transport.
\end{abstract}


\section{Introduction}
\label{sec:introduction}

Cosmic-ray electrons lose energy through synchrotron radiation and inverse-Compton scattering as they propagate through the Galaxy. For a loss rate $b(E)\simeq b_0E^2$ with $b_0\simeq10^{-16}~{\rm GeV^{-1}\,s^{-1}}$, the cooling time decreases from about $3\times10^5~{\rm yr}$ at $1~{\rm TeV}$ to $3\times10^4~{\rm yr}$ at $10~{\rm TeV}$. The observable volume therefore shrinks rapidly with energy. At sufficiently high energies the electron spectrum should no longer represent an average over a large Galactic source population, but it should instead become sensitive to a few recent accelerators in the Solar neighborhood~\citep{Shen1970Pulsars,Atoyan1995PRD,Aharonian1995AALetter,Kobayashi2004SNR,Delahaye2010Positrons}.

This connection between the loss horizon and individual sources has long motivated searches for structure in the TeV spectrum. Nearby supernova remnants and pulsar wind nebulae can produce broad excesses, changes in slope, or cutoffs whose positions depend on the source age, distance, and release time~\citep{Kobayashi2004SNR,Kawanaka2011Escape,Recchia2019LocalAccelerator}. In particular, \citet{Kobayashi2004SNR} showed that sources such as Vela, the Cygnus Loop, and Monogem may leave identifiable features in the TeV electron spectrum.

If escape from a young remnant is energy dependent, the spectrum can also acquire a low-energy boundary that carries information about when particles entered the interstellar medium~\citep{Kawanaka2011Escape}. More generally, the observed end of the electron spectrum may reflect acceleration, escape, propagation, source energetics, or detector sensitivity rather than a single universal cutoff~\citep{Sudoh2023ElectronEnd,JohnLinden2023NoSharpFeatures,Mertsch2018StochasticSources}.

Measurements by AMS-02, DAMPE, and CALET establish the all-electron spectrum through the TeV region and show a marked softening near $1~{\rm TeV}$~\citep{Aguilar2014AMSElectron,Aguilar2019AMSElectron,Ambrosi2017DAMPE,Adriani2023CALET,Guo2026DAMPE}. H.E.S.S. extends the electron-candidate spectrum to higher energies using atmospheric air showers~\citep{Aharonian2024HESS}, and dedicated searches for spectral and anisotropy signatures of Vela in the TeV electron data are under way with CALET~\citep{Motz2025CALETVela}. These measurements do not yet determine whether the spectrum above a few TeV is a smooth continuation or contains a contribution from a recent source. This is also the energy range in which a young nearby accelerator separates most clearly from the older and more distant source population. Extending direct measurements to higher energy therefore bears both on where the spectrum ends and on which process sets that end~\citep{Sudoh2023ElectronEnd,Adriani2025HERD}.

Vela provides a useful case study. Its distance is $287~{\rm pc}$ and its characteristic age is $11.3~{\rm kyr}$~\citep{Manchester2005ATNF}. Both the supernova remnant and the pulsar wind nebula contain electrons at multi-TeV energies, but the time profile of their escape into the interstellar medium remains uncertain. Particles may have been confined and released after the source evolved, while a longer-lived outflow can continue to supply young particles~\citep{Aharonian2006VelaX,Kawanaka2011Escape,DellaTorre2015VelaX,Huang2018VelaDiffusion}. These possibilities lead to different propagation times even when the injected spectrum and total released energy are the same.

Here we use Vela to examine how a young nearby source can shape the multi-TeV all-electron spectrum. We compare a birth-time burst with continuous release, scan the relevant diffusion scale, and separate the roles of the injection index $\alpha$ and the released electron energy $W_e$. We then extend the calculation to an axisymmetric diffusion tensor, for which the source visibility also depends on its orientation relative to the local ordered magnetic field. We do not attempt to fit a unique Vela realization to the present data; instead we identify spectral differences that can be tested as the measured energy range expands. This complements our earlier study of generic high-energy source features and stochastic inverse-Compton losses~\citep{XieYang2026Features} by focusing on the release history and local field geometry of a specific young source.

Section~\ref{sec:model} introduces the source and propagation model. Section~\ref{sec:isotropic} discusses the spectral signatures of release history, diffusion, and injection parameters. Section~\ref{sec:anisotropic} gives the corresponding mapping for anisotropic transport. The observational implications are discussed in Section~\ref{sec:prospects}, and Section~\ref{sec:conclusion} summarizes the conclusions.

\section{Source and propagation model}
\label{sec:model}

\subsection{Reference all-electron spectrum}

We write the all-electron flux as
\begin{equation}
J_{\rm all}(E)=J_{\rm bg}(E)+J_{\rm Vela}(E).
\label{eq:flux_decomposition}
\end{equation}
The reference background contains a continuous population of sources beyond $1~{\rm kpc}$ and burst-like components associated with Geminga and Monogem. The distant population follows the setup of \citet{HooperLinden2018Diffusion}, with
\begin{equation}
D_{\rm bg}(E)=3.86\times10^{28}
\left(\frac{E}{1~{\rm GeV}}\right)^{0.33}
{\rm cm^2\,s^{-1}}.
\label{eq:background_diffusion}
\end{equation}
The Geminga and Monogem spectral shapes are based on the single-source fits of \citet{Tang2022NearbySNR}. We retain their relative normalization and apply a common factor so that the composite background matches the published DAMPE all-electron flux at $200~{\rm GeV}$~\citep{Ambrosi2017DAMPE}. At this energy the Vela contribution in the representative model is small, so the procedure fixes the low-energy baseline without using the multi-TeV bins to determine the source. This construction supplies a reference scale against which the contrast of a young source can be assessed. This background is not fitted to the high-energy data, and the source-only results below do not depend on its normalization.

\subsection{Injection and release histories}

The spectrum released into the interstellar medium is parameterized as
\begin{equation}
q(E)=q_0
\left(\frac{E}{1~{\rm GeV}}\right)^{-\alpha}
\exp\left(-\frac{E}{E_{\rm cut}}\right),
\label{eq:injection}
\end{equation}
where the normalization is defined through the electron energy above $E_{\min}=1~{\rm GeV}$,
\begin{equation}
W_e=\int_{E_{\min}}^\infty dE\,E q(E).
\label{eq:source_energy}
\end{equation}
Unless stated otherwise, we use $\alpha=2.0$ and $E_{\rm cut}=100~{\rm TeV}$. The index is a conventional strong-shock reference value~\citep{Bell1978DSA,BlandfordOstriker1978DSA,Drury1983DSA}, while the cutoff is placed above the main energy range so that the propagated structure is not imposed a priori. Emission from young remnants and Vela X demonstrates that electrons can reach tens of TeV~\citep{Reynolds1999SNRElectrons,Aharonian2006VelaX}. We later vary $\alpha$ and $W_e$ independently.

Let $\tau$ be the propagation time, i.e.\ the time elapsed since release. The two limiting histories are
\begin{equation}
\begin{aligned}
{\cal R}_{\rm burst}(\tau;\tau_b)&=\delta(\tau-\tau_b),\\
{\cal R}_{\rm cont}(\tau)&=\frac{1}{T}\,
\Theta(\tau)\Theta(T-\tau),
\end{aligned}
\label{eq:release_histories}
\end{equation}
where $T=11.3~{\rm kyr}$. A birth-time burst has $\tau_b=T$. Delayed bursts with $\tau_b<T$ represent confinement followed by rapid release, whereas the continuous model spreads the same total energy uniformly over the source age. These profiles are deliberately simple. These two profiles represent limiting cases of the propagation-time distribution.

\subsection{Diffusion-loss solution}

The electron density obeys
\begin{equation}
\frac{\partial\psi}{\partial t}
-\bm{\nabla}\cdot\left(\bm{D}\bm{\nabla}\psi\right)
-\frac{\partial}{\partial E}\left[b(E)\psi\right]
=Q(E,t)\delta^3(\bm{x}-\bm{x}_s).
\label{eq:transport}
\end{equation}
We first assume isotropic diffusion,
\begin{equation}
D(E)=D_{\rm 1TeV}
\left(\frac{E}{1~{\rm TeV}}\right)^{\delta},
\qquad \delta=\frac{1}{3},
\label{eq:isotropic_diffusion}
\end{equation}
and continuous radiative losses,
\begin{equation}
b(E)=b_0E^2,
\qquad b_0=10^{-16}~{\rm GeV^{-1}\,s^{-1}}.
\label{eq:continuous_loss}
\end{equation}
For an observed energy $E$ and propagation time $\tau$, the injection energy is
\begin{equation}
E_s=\frac{E}{1-b_0E\tau}.
\label{eq:source_energy_mapping}
\end{equation}
The diffusion-loss length and the isotropic spatial kernel are
\begin{equation}
\begin{aligned}
\lambda^2(E,E_s)&=4\int_E^{E_s}dE'\frac{D(E')}{b(E')},\\
G_{\rm iso}(\lambda,d)&=
\frac{1}{(\pi\lambda^2)^{3/2}}
\exp\left(-\frac{d^2}{\lambda^2}\right).
\end{aligned}
\label{eq:isotropic_green}
\end{equation}
The present density for either history is
\begin{equation}
\psi(E,d,T)=\int_0^{\tau_{\max}}d\tau\,
{\cal R}(\tau)\frac{b(E_s)}{b(E)}q(E_s)
G_{\rm iso}\!\left[\lambda(E,E_s),d\right],
\label{eq:history_solution}
\end{equation}
where $\tau_{\max}=\min[T,1/(b_0E)]$, and $J(E)=c\psi(E)/(4\pi)$~\citep{Atoyan1995PRD,Delahaye2010Positrons}. Only the transport of the Vela component is varied; the reference background is kept fixed.

Table~\ref{tab:model_parameters} collects the parameters used below. Figure~\ref{fig:release_spectra} shows the transport trend across the scanned diffusion range. The $(W_e,\alpha)$ map and the representative spectra use the stated benchmark diffusion coefficient, so that changes in injection and release history can be compared without a simultaneous change in transport.

\begin{table}[t]
\centering
\caption{Source and transport parameters. Ranges denote quantities varied in the corresponding scan.}
\label{tab:model_parameters}
\renewcommand{\arraystretch}{1.12}
\begin{tabular}{lcc}
\hline
Quantity & Value or range & Use \\
\hline
$d$ & $287~{\rm pc}$ & fixed \\
$T$ & $11.3~{\rm kyr}$ & fixed \\
$b_0$ & $10^{-16}~{\rm GeV^{-1}\,s^{-1}}$ & fixed \\
$\delta$ & $1/3$ & fixed \\
$E_{\rm cut}$ & $100~{\rm TeV}$ & fixed \\
$D_{\rm 1TeV}$ & $3\times10^{28}$--$3\times10^{30}~{\rm cm^2\,s^{-1}}$ & Fig.~\ref{fig:release_spectra} \\
$D_{\rm 1TeV}$ & $5\times10^{29}~{\rm cm^2\,s^{-1}}$ & benchmark \\
$\alpha$ & $1.5$--$2.6$ & Fig.~\ref{fig:energy_index} \\
$W_e(>1~{\rm GeV})$ & $10^{44}$--$10^{48}~{\rm erg}$ & Fig.~\ref{fig:energy_index} \\
\hline
\end{tabular}
\end{table}

\subsection{Spectral observables}

Instead of defining a normalization from an individual data bin, we characterize a model by the shape and contrast of its source contribution. The fractional Vela contribution is
\begin{equation}
f_{\rm Vela}(E)=
\frac{J_{\rm Vela}(E)}{J_{\rm bg}(E)+J_{\rm Vela}(E)}.
\label{eq:source_fraction}
\end{equation}
For the parameter scan we use the Vela fraction at $5~{\rm TeV}$,
\begin{equation}
f_5\equiv f_{\rm Vela}(5~{\rm TeV}),
\label{eq:f5}
\end{equation}
and denote by $E_{\rm feature}$ the energy at which $E^3J_{\rm Vela}$ is largest. This pivot energy ($5~{\rm TeV}$) lies just beyond the range in which present direct measurements have useful statistics, while remaining low enough to avoid the trivial rise of the source fraction in the far tail, where the reference background falls rapidly. The quantities have separate roles: $E_{\rm feature}$ follows the propagation and cooling scales, whereas $f_5$ measures the contrast that an extended direct measurement would need to resolve.

For fixed $\alpha$, the Vela flux is linear in $W_e$. The normalization of Eq.~(\ref{eq:injection}) can be written as
\begin{equation}
\begin{aligned}
q_0&=\frac{W_e}{I(\alpha,E_{\rm cut})},\\
I&=\int_{E_{\min}}^\infty dE\,E
\left(\frac{E}{1~{\rm GeV}}\right)^{-\alpha}
e^{-E/E_{\rm cut}}.
\end{aligned}
\label{eq:injection_normalization}
\end{equation}
Working in the $(W_e,\alpha)$ plane separates two effects that are otherwise degenerate over a narrow energy interval: $W_e$ sets the overall normalization and energy budget of the source contribution, while $\alpha$ changes the fraction of that energy released at multi-TeV energies. A measurement that covers both the rise and turnover of the source contribution can distinguish them more effectively than a single flux point.

\section{Spectral signatures under isotropic diffusion}
\label{sec:isotropic}

\begin{figure*}
\centering
\includegraphics[width=0.98\textwidth]{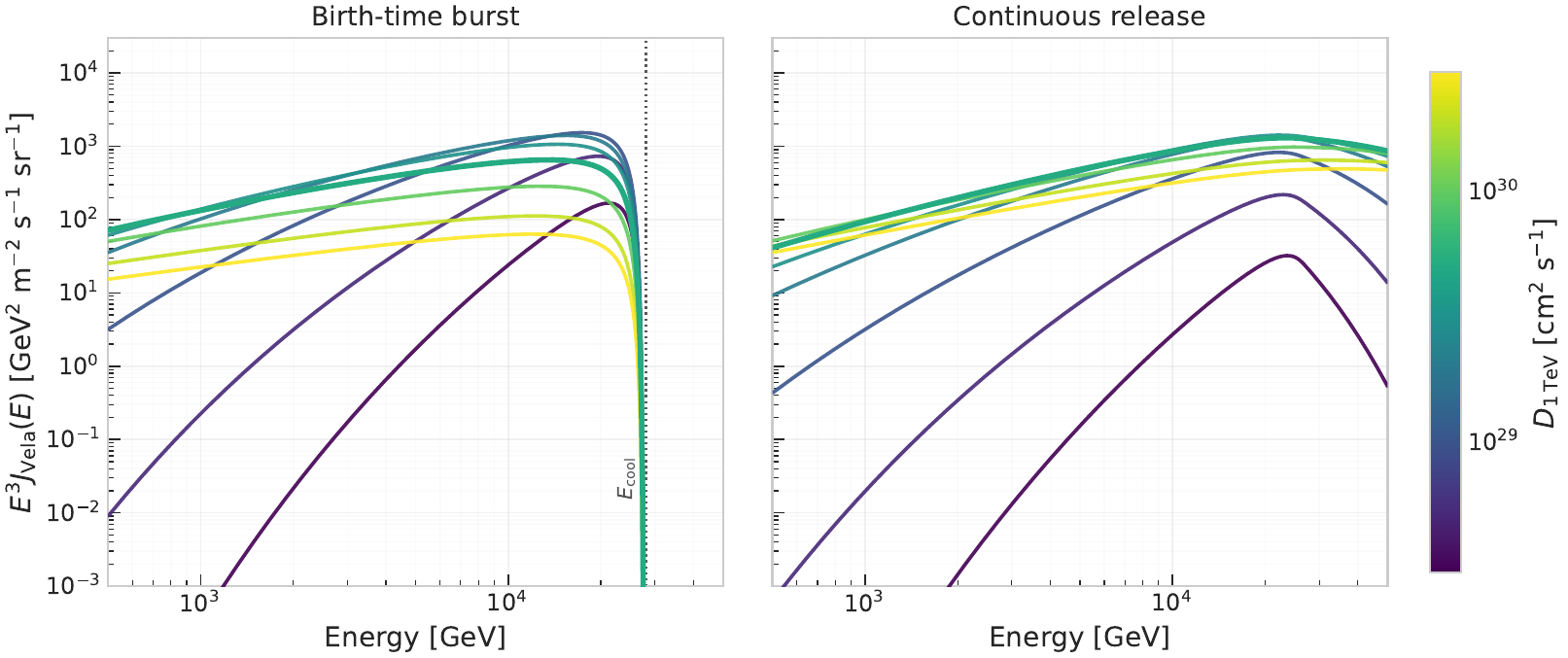}
\caption{Vela source spectra for birth-time burst release (left) and continuous release (right) under isotropic diffusion. The injection spectrum and released electron energy are identical in both panels: $\alpha=2.0$, $E_{\rm cut}=100~{\rm TeV}$, and $W_e=10^{48}~{\rm erg}$. The color sequence spans nine values of $D_{\rm 1TeV}$ from $3\times10^{28}$ to $3\times10^{30}~{\rm cm^2\,s^{-1}}$; the $5\times10^{29}~{\rm cm^2\,s^{-1}}$ curve is slightly thicker. The dotted line marks the age-related cooling boundary of the birth-time burst. }
\label{fig:release_spectra}
\end{figure*}

Figure~\ref{fig:release_spectra} compares the Vela source spectra for the two release histories across two decades in $D_{\rm 1TeV}$. The curves share $W_e=10^{48}~{\rm erg}$, $\alpha=2.0$, and $E_{\rm cut}=100~{\rm TeV}$, so the changes in shape trace transport and the distribution of release times.

In the birth-time burst, every electron propagates for the source age $T$. Equation~(\ref{eq:source_energy_mapping}) then sets the cooling boundary
\begin{equation}
E_{\rm cool}=\frac{1}{b_0T}\simeq28~{\rm TeV}.
\label{eq:cooling_boundary}
\end{equation}
The burst spectrum falls rapidly near this energy. Continuous release contains particles with propagation times between zero and $T$. Its youngest electrons retain more of their injection energy and extend the spectrum above $E_{\rm cool}$. The contrast between the two panels therefore arises from the release-time distribution, with the same injection cutoff in both cases.

The diffusion coefficient controls both arrival and dilution. For $\lambda\ll d$, Earth lies outside most of the diffusion cloud and receives few particles. For $\lambda\gg d$, the particles fill a large volume and the local density drops. At a fixed propagation time, the kernel in Eq.~(\ref{eq:isotropic_green}) reaches its maximum at
\begin{equation}
\lambda^2=\frac{2d^2}{3}.
\label{eq:optimal_length}
\end{equation}
A burst samples this balance at one propagation time, whereas continuous release superposes particles with $0<\tau<T$. Its shape reflects contributions from both sides of the efficient propagation scale.

The source contrast is then set by the released energy $W_e$ and injection index $\alpha$. Increasing $W_e$ raises the spectrum at all energies. Decreasing $\alpha$ places a larger share of the energy budget in multi-TeV electrons. Lines of fixed contrast therefore run diagonally across the $(W_e,\alpha)$ plane: softer injection requires more released energy to give the same high-energy flux.

\begin{figure*}
\centering
\includegraphics[width=0.96\textwidth]{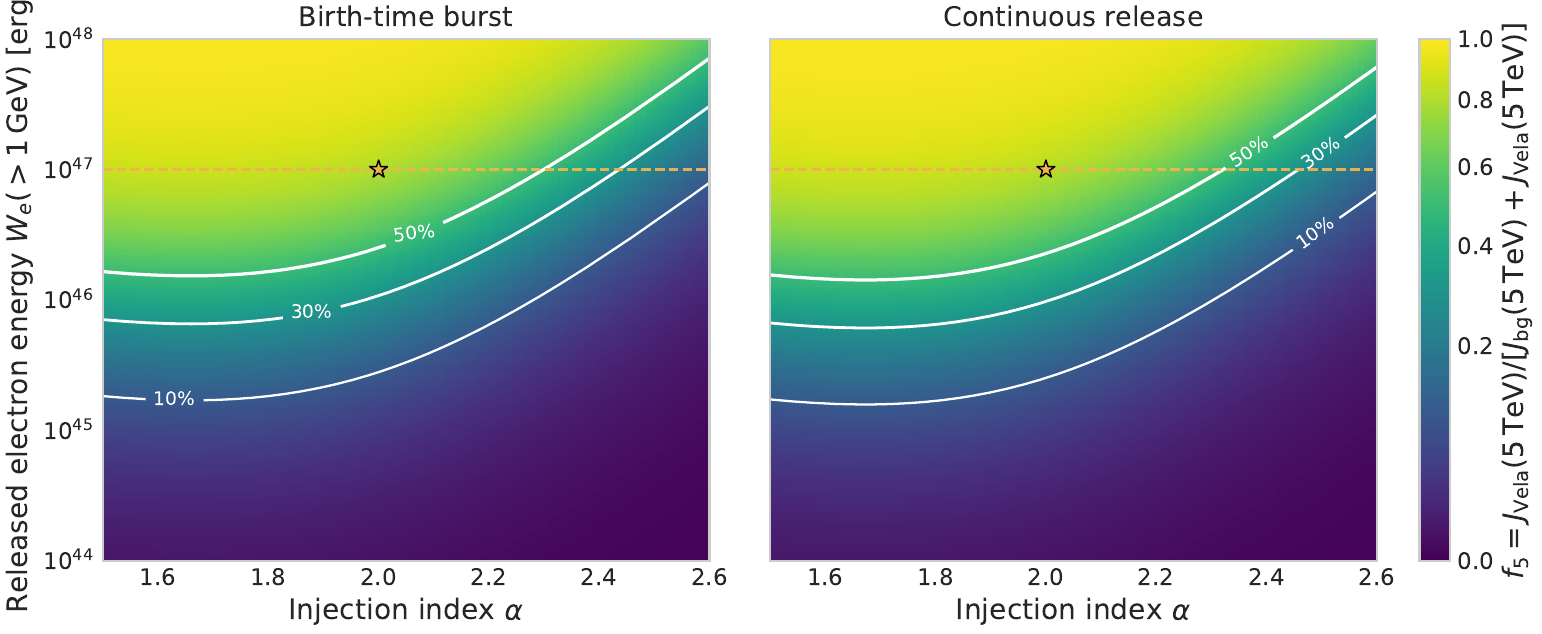}
\caption{Predicted Vela fraction at $5~{\rm TeV}$ in the plane of released electron energy and injection index for birth-time burst (left) and continuous release (right). The contours mark $f_5=0.1$, 0.3, and 0.5. The star denotes the high-contrast illustrative model in Fig.~\ref{fig:future_spectra}, with $\alpha=2.0$ and $W_e=10^{47}~{\rm erg}$. The map gives spectral predictions at $D_{\rm 1TeV}=5\times10^{29}~{\rm cm^2\,s^{-1}}$; it is not an allowed region derived from current data.}
\label{fig:energy_index}
\end{figure*}

Figure~\ref{fig:energy_index} shows this relation through $f_5$ at $D_{\rm 1TeV}=5\times10^{29}~{\rm cm^2\,s^{-1}}$. The 10\%, 30\%, and 50\% contours are model thresholds. The contours are parallel because $f_5$ is a fixed monotone function of $J_{\rm Vela}/J_{\rm bg}$, and $J_{\rm Vela}\propto W_e$ at fixed $\alpha$. At $5~{\rm TeV}$, the cooling time exceeds the Vela age, so the burst and continuous-release contours remain close. The release histories separate near the burst cooling boundary.

\begin{samepage}
Table~\ref{tab:energy_index_thresholds} lists representative points on the 10\% and 30\% contours. For $\alpha=2.0$, a 10\% contribution requires $W_e=2.80\times10^{45}~{\rm erg}$ for a burst and $2.52\times10^{45}~{\rm erg}$ for continuous release. The fraction of injected energy above $1~{\rm TeV}$ falls from $36.9\%$ at $\alpha=2.0$ to $4.94\%$ at $\alpha=2.4$, so harder spectra require proportionally more released energy (Table~\ref{tab:energy_index_thresholds}). The release histories differ little at the $5~{\rm TeV}$ pivot.
\end{samepage}

\begin{table*}[t]
\centering
\caption{Released energy required to obtain a specified Vela fraction at $5~{\rm TeV}$. Energies are in units of $10^{45}~{\rm erg}$ and refer to electrons above $1~{\rm GeV}$. The column $F_{>1\,\mathrm{TeV}}$ is the fraction of injected electron energy above $1~{\rm TeV}$.}
\label{tab:energy_index_thresholds}
\renewcommand{\arraystretch}{1.13}
\begin{tabular}{cccccc}
\hline
& & \multicolumn{2}{c}{Birth-time burst} & \multicolumn{2}{c}{Continuous release} \\
\cline{3-6}
$\alpha$ & $F_{>1\,\mathrm{TeV}}$ & $W_e(f_5=0.1)$ & $W_e(f_5=0.3)$ & $W_e(f_5=0.1)$ & $W_e(f_5=0.3)$ \\
\hline
1.6 & 0.831 & 1.72 & 6.63 & 1.61 & 6.20 \\
1.8 & 0.636 & 1.83 & 7.06 & 1.68 & 6.49 \\
2.0 & 0.369 & 2.80 & 10.8 & 2.52 & 9.73 \\
2.2 & 0.153 & 6.45 & 24.9 & 5.72 & 22.1 \\
2.4 & 0.0494 & 20.5 & 79.2 & 17.9 & 69.0 \\
\hline
\end{tabular}
\end{table*}

The position of the feature provides information that its amplitude alone does not. As $\alpha$ changes from 1.6 to 2.4, the maximum of $E^3J_{\rm Vela}$ moves from $18.3$ to $6.69~{\rm TeV}$ for the birth-time burst and from $29.8$ to $15.6~{\rm TeV}$ for continuous release. A hard source with a modest energy budget and a softer source with a larger energy budget can share the same $f_5$ while producing different slopes and turnover energies. Measuring the rise and decline of the feature would constrain $W_e$, $\alpha$, and the release history together.

\begin{figure*}
\centering
\includegraphics[width=0.6\textwidth]{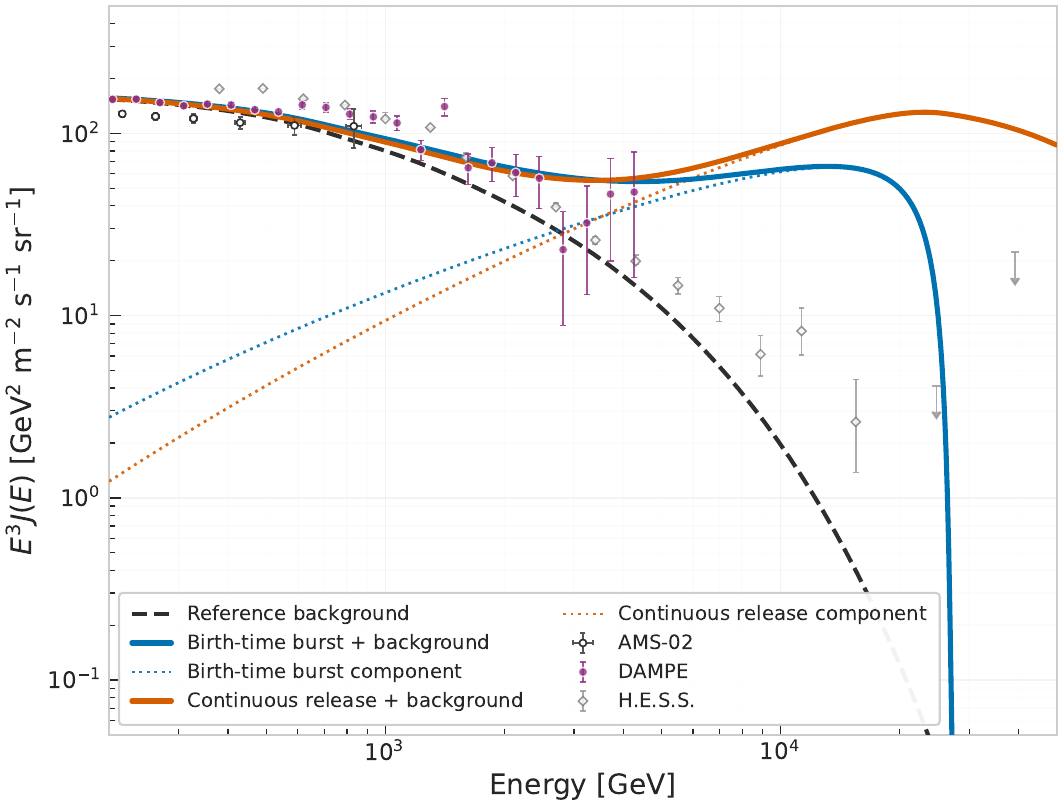}
\caption{Illustrative all-electron spectra for Vela with birth-time burst or continuous release. Both models use $\alpha=2.0$, $W_e=10^{47}~{\rm erg}$, $E_{\rm cut}=100~{\rm TeV}$, and $D_{\rm 1TeV}=5\times10^{29}~{\rm cm^2\,s^{-1}}$. Solid curves include the reference background; dotted curves show the Vela components. AMS-02, DAMPE, and H.E.S.S. data are shown for comparison, with the two highest-energy H.E.S.S. points displayed as upper limits. The background is anchored only to the published DAMPE flux at $200~{\rm GeV}$; neither the Vela curves nor the background are fitted to H.E.S.S. The common source normalization is a rounded illustrative benchmark rather than a quoted best fit.}
\label{fig:future_spectra}
\end{figure*}

Figure~\ref{fig:future_spectra} places these source spectra on the observational scale by adding the reference background. Both release histories use $\alpha=2.0$, $W_e=10^{47}~{\rm erg}$, and $D_{\rm 1TeV}=5\times10^{29}~{\rm cm^2\,s^{-1}}$. This rounded source energy gives Vela fractions of 79.9\% for the burst and 81.5\% for continuous release at $5~{\rm TeV}$, and serves as an illustrative high-contrast case. AMS-02 and DAMPE set the scale of the direct measurements, while H.E.S.S. provides an indirect high-energy comparison. Above the current direct range, the burst component peaks near $13.5~{\rm TeV}$ and ends at its cooling boundary. The continuous component peaks near $23.1~{\rm TeV}$ and forms a broader tail.

\begin{figure*}
\centering
\includegraphics[width=0.98\textwidth]{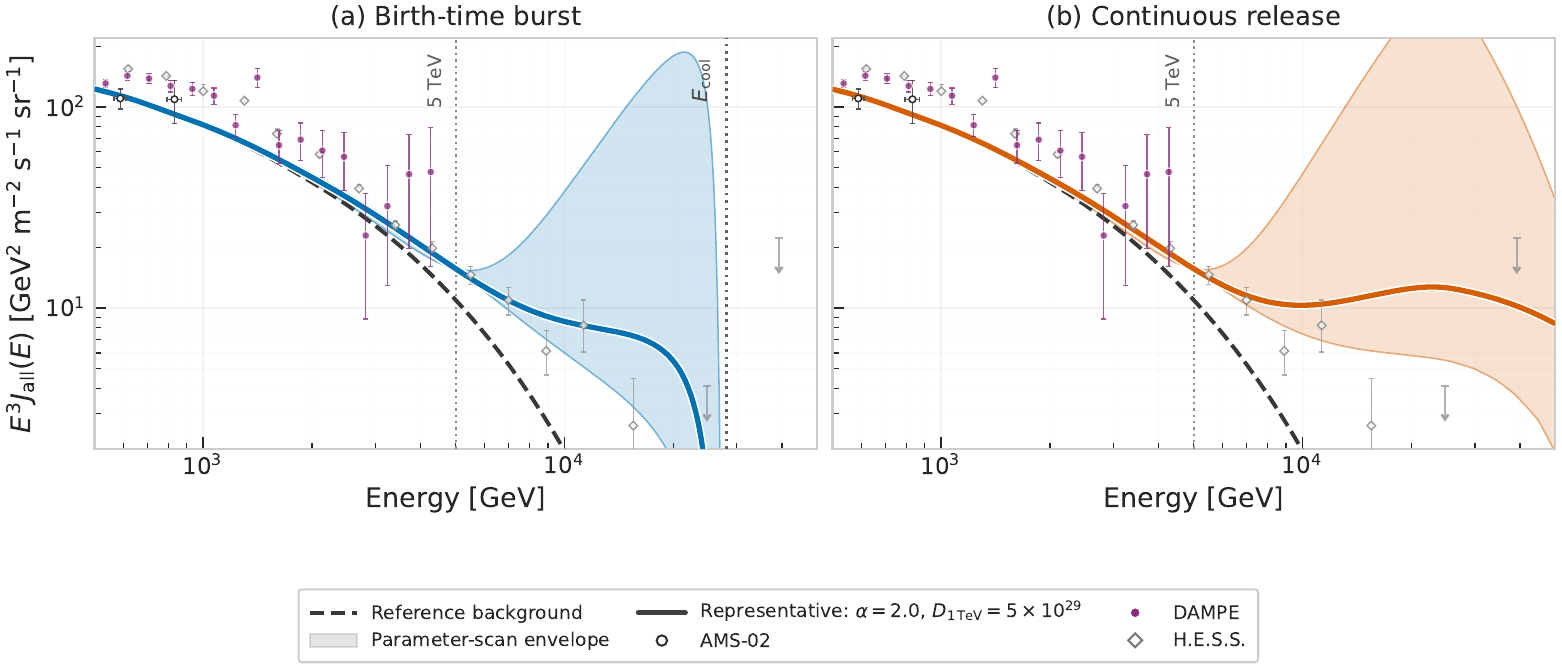}
\caption{Expected all-electron spectral shapes at fixed $f_5=0.30$ for a birth-time burst (left) and continuous release (right). The shaded regions span $\alpha=1.6$--2.4 and $D_{\rm 1TeV}=3\times10^{28}$--$3\times10^{30}~{\rm cm^2\,s^{-1}}$, after adjusting $W_e$ separately for every model and requiring $W_e\leq10^{48}~{\rm erg}$. Thick curves show $\alpha=2.0$ and $D_{\rm 1TeV}=5\times10^{29}~{\rm cm^2\,s^{-1}}$; the dashed curve is the reference background. AMS-02, DAMPE, and H.E.S.S. are shown only as observational context, and the two highest-energy H.E.S.S. measurements are plotted as upper limits. The vertical line at $5~{\rm TeV}$ marks the common normalization energy, while $E_{\rm cool}$ marks the age-related burst boundary. }
\label{fig:spectral_envelopes}
\end{figure*}

The fixed $W_e$ in Fig.~\ref{fig:future_spectra} leaves a small difference in source contrast between the two release histories. In Fig.~\ref{fig:spectral_envelopes} we remove that difference by fixing $f_5=0.30$ in each model. We scan $\alpha=1.6$--2.4 and $D_{\rm 1TeV}=3\times10^{28}$--$3\times10^{30}~{\rm cm^2\,s^{-1}}$, adjusting $W_e$ at every point. Models requiring $W_e>10^{48}~{\rm erg}$ are excluded. The shaded regions collect the predicted spectra from this scan; they do not represent confidence intervals or data-selected regions.

For the birth-time burst, the common propagation time produces a maximum below $E_{\rm cool}\simeq1/(b_0T)$ and a rapid decline toward the cooling boundary. The representative curve peaks at $13.5~{\rm TeV}$ and requires $W_e=1.08\times10^{46}~{\rm erg}$. The abrupt terminal edge follows from the continuous-loss approximation; stochastic inverse-Compton losses would broaden it. Continuous release adds young electrons with short cooling times, producing a wider feature that crosses $E_{\rm cool}$. Its representative curve peaks at $23.1~{\rm TeV}$ and requires $W_e=9.73\times10^{45}~{\rm erg}$.

AMS-02 and DAMPE stop short of the maxima of these predicted features and do not sample their declining sides. H.E.S.S. reaches this energy range but is included only as an indirect comparison. A direct spectrum with a peak near $10~{\rm TeV}$ followed by a rapid decline would favor an early burst. A broad tail extending to several tens of TeV would favor continuous release. The shape across the feature carries this distinction; a single high-energy bin cannot provide it.

\section{Anisotropic diffusion and field geometry}\label{sec:anisotropic}

The isotropic calculation contains no information about the direction of the local magnetic field. To retain this geometry, we use an axisymmetric diffusion tensor~\citep{Effenberger2012Anisotropic,NavaGabici2013,Giacinti2018Anisotropic},
\begin{equation}
\bm{D}=D_\perp\bm{I}
+(D_\parallel-D_\perp)\hat{\bm b}\hat{\bm b},
\qquad \eta\equiv\frac{D_\perp}{D_\parallel},
\label{eq:diffusion_tensor}
\end{equation}
where $\hat{\bm b}$ is the effective ordered-field direction. We take $D_\parallel\propto E^{1/3}$ and an energy-independent ratio $\eta$. If $\theta$ is the angle between the Vela--Earth displacement vector and $\hat{\bm b}$, the spatial kernel becomes
\begin{equation}
G_{\rm ani}=
\frac{1}{\eta(\pi\lambda_\parallel^2)^{3/2}}
\exp\left(-\frac{d_{\rm eff}^2}{\lambda_\parallel^2}\right),
\label{eq:anisotropic_green}
\end{equation}
with
\begin{equation}
d_{\rm eff}=d
\left(\cos^2\theta+\frac{\sin^2\theta}{\eta}\right)^{1/2}.
\label{eq:effective_distance}
\end{equation}
For $\theta=0$, propagation proceeds along the ordered field and the transverse confinement of the particles raises the local density. For $\theta>0$, the perpendicular part of the source displacement enters $d_{\rm eff}^2$ with the factor $1/\eta$, so a small cross-field diffusion coefficient can strongly suppress the Vela flux.

For an energy-independent $\eta$ and the same energy dependence of $D_\parallel$ and $D_\perp$, the anisotropic result follows directly from the isotropic solution. Defining
\begin{equation}
F(\eta,\theta)=\cos^2\theta+\frac{\sin^2\theta}{\eta},
\end{equation}
we obtain
\begin{equation}
J_{\rm ani}(D_\parallel,\eta,\theta)
=\frac{J_{\rm iso}(D_\parallel/F)}{\eta F^{3/2}}.
\label{eq:anisotropic_rescaling}
\end{equation}
Equation~(\ref{eq:anisotropic_rescaling}) maps the isotropic spectrum to a chosen $(D_\parallel,\eta,\theta)$ without repeating the time integration. The angle changes both the effective propagation scale and the normalization. At fixed $W_e$, alignment enhances the Vela contribution, whereas a displacement from the field direction suppresses it. Since $W_e$, $\eta$, and $\theta$ all affect the source amplitude, the electron spectrum alone cannot determine the local diffusion tensor~\citep{Liu2019GemingaAnisotropic,Xia2025GemingaAnisotropic}.

\begin{figure*}
\centering
\includegraphics[width=0.92\textwidth]{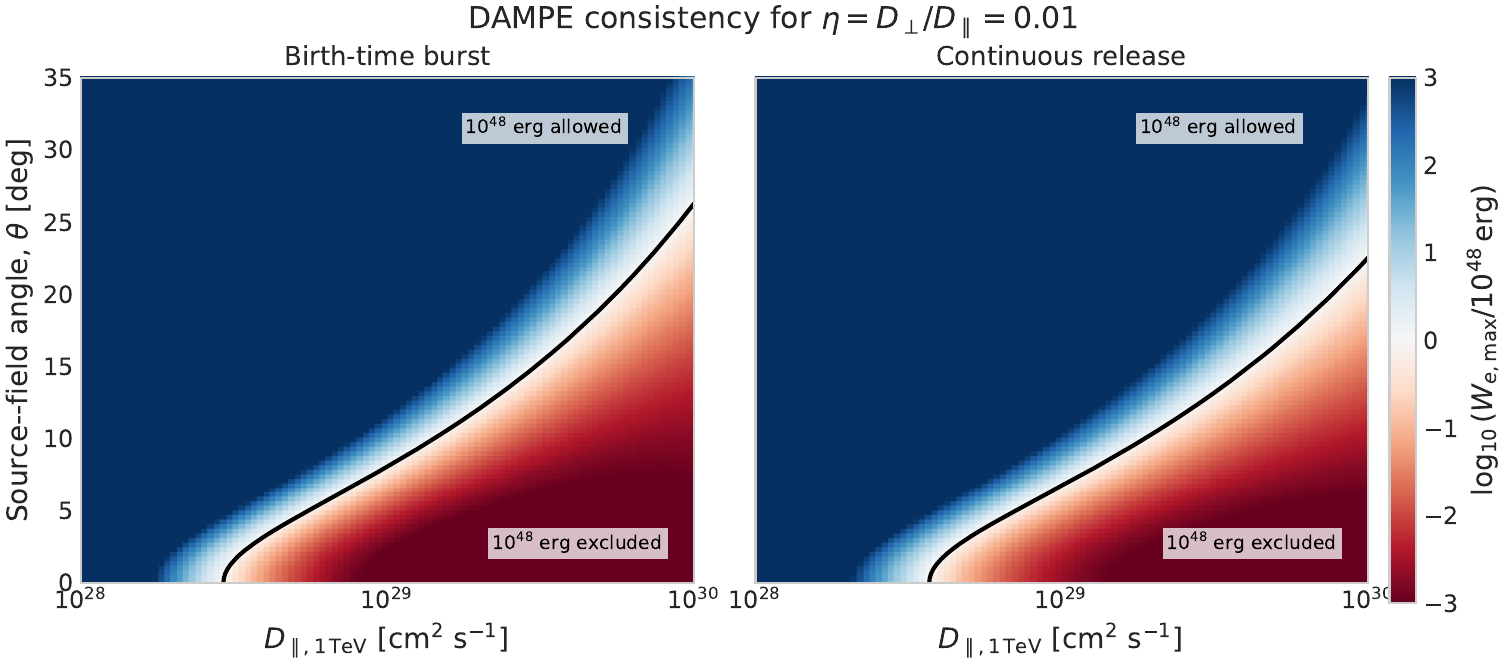}
\caption{Maximum released electron energy compatible with the adopted DAMPE upper envelope in the $(D_{\parallel,\rm 1TeV},\theta)$ plane, for a birth-time burst (left) and continuous release (right) with $\eta=0.01$. The color scale gives $\log_{10}(W_{e,\max}/10^{48}~{\rm erg})$, and the black contour marks $W_{e,\max}=10^{48}~{\rm erg}$. For that fixed source energy, the blue side of the contour is compatible with the adopted envelope and the red side exceeds it. The result is conditional on the reference background.}
\label{fig:anisotropic_bounds}
\end{figure*}

We illustrate the extent of this degeneracy with a conditional comparison to DAMPE. At each point in the $(D_{\parallel,\rm 1TeV},\theta)$ plane, the Vela normalization is increased until the total spectrum first reaches the one-sigma upper envelope of the DAMPE bins above $1~{\rm TeV}$. Figure~\ref{fig:anisotropic_bounds} gives the corresponding maximum released energy $W_{e,\max}$ for $\eta=0.01$. The black contour is $W_{e,\max}=10^{48}~{\rm erg}$: a source with this energy lies below the adopted envelope where $W_{e,\max}>10^{48}~{\rm erg}$ (blue) and exceeds it where $W_{e,\max}<10^{48}~{\rm erg}$ (red). This construction exposes the normalization--geometry degeneracy; it is not a likelihood fit for the field direction.

The black contour defines the minimum angle listed in Table~\ref{tab:critical_angles}. At $D_{\parallel,\rm 1TeV}=10^{28}~{\rm cm^2\,s^{-1}}$, even an aligned source with $W_e=10^{48}~{\rm erg}$ remains below the adopted envelope because the particles arrive inefficiently. The minimum angle grows with $D_{\parallel,\rm 1TeV}$ over the range shown, reaching $26.3^\circ$ for a burst and $22.5^\circ$ for continuous release at $10^{30}~{\rm cm^2\,s^{-1}}$. These angles are consistency thresholds within a uniform-tensor model, not estimates of the actual field direction. Field-line wandering, variation of the mean field over the $287~{\rm pc}$ source distance, and a possible slow-diffusion region around Vela are not included~\citep{Giacinti2013Gamma,LopezCoto2018Turbulence,Huang2018VelaDiffusion,Abeysekara2017Geminga,Wu2026AnisoHalo,Yan2025Coherence}.

\begin{table}[ht]
\centering
\caption{Minimum source--field angle for which a model with $\eta=0.01$ and $W_e=10^{48}~{\rm erg}$ remains below the adopted DAMPE upper envelope.}
\label{tab:critical_angles}
\renewcommand{\arraystretch}{1.15}
\begin{tabular}{ccc}
\hline
$D_{\parallel,\rm 1TeV}$ [$\mathrm{cm^2\,s^{-1}}$]
& $\theta_{\min}^{\rm burst}$ & $\theta_{\min}^{\rm cont}$ \\
\hline
$1\times10^{28}$ & $0.00^\circ$ & $0.00^\circ$ \\
$5\times10^{28}$ & $4.50^\circ$ & $3.16^\circ$ \\
$1\times10^{29}$ & $8.06^\circ$ & $6.72^\circ$ \\
$5\times10^{29}$ & $19.1^\circ$ & $16.5^\circ$ \\
$1\times10^{30}$ & $26.3^\circ$ & $22.5^\circ$ \\
\hline
\end{tabular}
\end{table}

Radio observations offer an independent route to constrain the field geometry entering our transport model. For a pulsar, the rotation and dispersion measures give the electron-density-weighted line-of-sight field through $\langle B_\parallel\rangle=1.232\,{\rm RM}/{\rm DM}~\mu{\rm G}$, with RM in rad m$^{-2}$ and DM in pc cm$^{-3}$. Distance-resolved pulsar samples have been used to reconstruct the large-scale Galactic magnetic field~\citep{Xu2022FASTMagnetic,SetaFederrath2021}. Under the uniform-field approximation adopted here, an estimate of the ordered-field strength would relate this quantity to the source--field angle through $|\cos\theta|\simeq|\langle B_\parallel\rangle|/B_{\rm ord}$. Measurements toward Vela could therefore provide an external prior on $\theta$. The RM/DM estimator may be biased on sub-kpc scales when the thermal electron density is correlated with small-scale magnetic structure~\citep{SetaFederrath2021}. Moreover, the RM and DM of the Vela pulsar vary on long timescales because its line of sight passes through magnetized structures associated with the Vela supernova remnant and the Gum Nebula~\citep{Lee2024VelaRM}. We therefore retain $\theta$ as an effective parameter. Future RM/DM tomography of pulsars at different distances in the Vela region, combined with polarization measurements of the projected field, could test the angular thresholds reported in Table~\ref{tab:critical_angles}.

\section{Observational prospects and limitations}
\label{sec:prospects}

The discriminating observable is how the spectrum evolves across the source feature, not its flux at a single energy. In the burst model, the rise, maximum, and cooling edge are tied to one propagation time. Continuous release superposes particles of different ages and broadens the high-energy tail. Figure~\ref{fig:spectral_envelopes} shows that the present direct measurements end below the range in which these shapes separate most clearly. The $(W_e,\alpha)$ map then relates the contrast of a future feature to the source energy and injection index.

Space-based calorimeters and ground-based air-shower instruments have different roles in this problem. DAMPE and CALET directly measure charged particles with calorimetric energy reconstruction and particle identification~\citep{Guo2026DAMPE,Adriani2023CALET}. H.E.S.S. gains a much larger collection area but infers an electron-candidate spectrum from atmospheric showers after statistical subtraction of a hadronic background. The H.E.S.S. analysis reports that residual proton contamination becomes increasingly important above a few TeV and treats its two highest-energy bins as upper limits. Its atmospheric, interaction-model, and energy-scale systematics also differ from those of a satellite calorimeter~\citep{Aharonian2024HESS}. We therefore show H.E.S.S. in Figs.~\ref{fig:future_spectra} and~\ref{fig:spectral_envelopes} only as an indirect comparison. It is not used to normalize the background, select a Vela model, or define the $(W_e,\alpha)$ contours.

The decisive test is an extension of the direct all-electron measurement. Higher DAMPE and CALET exposure can reduce the statistical uncertainty in the last direct bins~\citep{Guo2026DAMPE}, while a next-generation calorimeter such as HERD is designed to combine a large acceptance with electron--hadron separation~\citep{Adriani2023CALET,Sudoh2023ElectronEnd,Adriani2025HERD}. Coverage through $5$--$30~{\rm TeV}$ would test both parts of the prediction: the growth of a nearby-source contribution and the different spectral turnovers associated with burst and continuous release. Ground-based data remain a useful independent comparison, but an instrument-specific joint likelihood is outside the scope of this work.

Several model limitations remain. The reference background is phenomenological and its uncertainty propagates into the Vela fraction. The release histories represent two limiting cases but do not include energy-dependent escape or pulsar spin-down. The continuous-loss approximation neglects fluctuations in inverse-Compton scattering, which can modify a sharp high-energy edge; that effect was considered separately in our previous work~\citep{XieYang2026Features}.

The uniform diffusion tensor is an effective description and does not resolve the magnetic structure between Vela and the solar system. Any quantitative inference of source parameters will need to account for these uncertainties once a spectral feature is established over a sufficiently broad energy range.

\section{Conclusion}
\label{sec:conclusion}

The multi-TeV all-electron spectrum is a probe of the local source population because radiative losses progressively remove old and distant accelerators. Using Vela as a representative young nearby system, we find that the release history leaves a direct imprint on the high-energy shape. A birth-time burst develops an age-related cooling boundary near $1/(b_0T)$, whereas the continuous-release spectrum contains recently escaped particles and extends to higher energies. The source flux is maximized over a finite diffusion range, bounded by inefficient arrival at small diffusion coefficients and strong spatial dilution at large diffusion coefficients.

The released energy and injection index form a natural parameter plane for future tests. At $D_{\rm 1TeV}=5\times10^{29}~{\rm cm^2\,s^{-1}}$ and $\alpha=2.0$, a 10\% contribution at $5~{\rm TeV}$ requires $W_e=2.80\times10^{45}~{\rm erg}$ for a birth-time burst and $2.52\times10^{45}~{\rm erg}$ for continuous release. For $\alpha=2.4$, the required energies are larger by about a factor of seven. Measurements that span the rise and turnover can break this energy--index degeneracy and distinguish release histories more reliably than a normalization based on the last measured bin. Anisotropic diffusion adds a geometric dependence: alignment with the ordered field can enhance the source, whereas an offset from that direction can suppress it. The electron spectrum is therefore a joint probe of source release and local transport, and it constrains the source energetics and the local field geometry at the same time.

The present direct data define the lower-energy baseline but do not require a unique Vela component. Direct all-electron measurements extended through the $5$--$30~{\rm TeV}$ range, compared with independent ground-based data and with information on the local magnetic field, would show whether a young nearby accelerator shapes the end of the Galactic electron spectrum. Within the release histories studied here, a peak followed by a rapid decline would indicate an early burst, whereas a broad tail extending to several tens of TeV would favor continuous release.

\begin{acknowledgments}
Ruizhi Yang is supported by the NSFC under grants 12588101 and 12393854, and by the Natural Science Foundation of Sichuan Province under grant 2025ZNSFSC0065. Ruizhi Yang acknowledges support from the Cyrus Chun Ying Tang Foundations and the studio of Academician Zhao Zhengguo, Deep Space Exploration Laboratory.
\end{acknowledgments}

\bibliographystyle{astronomycomm}
\bibliography{main}

\end{document}